\documentclass[aps,prl,reprint,superscriptaddress]{revtex4-2}
\usepackage{amsmath}
\usepackage{amssymb}
\usepackage{mathtools}
\usepackage{xcolor}
\usepackage{graphicx}
\usepackage[normalem]{ulem}
\usepackage{physics}
\usepackage{braket}

\begin{document}

\title{Non-Gaussian state generation with time-gated photon detection}


\author{Tatsuki Sonoyama}
\email{sonoyama@alice.t.u-tokyo.ac.jp}
\affiliation{Department of Applied Physics, School of Engineering, The University of Tokyo, \\ 7-3-1 Hongo, Bunkyo-ku, Tokyo 113-8656, Japan}
\author{Kazuma Takahashi}
\affiliation{Department of Applied Physics, School of Engineering, The University of Tokyo, \\ 7-3-1 Hongo, Bunkyo-ku, Tokyo 113-8656, Japan}
\author{Baramee Charoensombutamon}
\affiliation{Department of Applied Physics, School of Engineering, The University of Tokyo, \\ 7-3-1 Hongo, Bunkyo-ku, Tokyo 113-8656, Japan}
\author{\\Sachiko Takasu}
\affiliation{National Institute of Advanced Industrial Science and Technology, \\ Tsukuba, Ibaraki, 305-8563, Japan}
\author{Kaori Hattori}
\affiliation{National Institute of Advanced Industrial Science and Technology, \\ Tsukuba, Ibaraki, 305-8563, Japan}
\affiliation{AIST-UTokyo Advanced Operando-Measurement Technology Open Innovation Laboratory, \\ Tsukuba, Ibaraki, 305-8563, Japan}
\author{Daiji Fukuda}
\affiliation{National Institute of Advanced Industrial Science and Technology, \\ Tsukuba, Ibaraki, 305-8563, Japan}
\affiliation{AIST-UTokyo Advanced Operando-Measurement Technology Open Innovation Laboratory, \\ Tsukuba, Ibaraki, 305-8563, Japan}
\author{Kosuke Fukui}
\affiliation{Department of Applied Physics, School of Engineering, The University of Tokyo, \\ 7-3-1 Hongo, Bunkyo-ku, Tokyo 113-8656, Japan}
\author{Kan Takase}
\affiliation{Department of Applied Physics, School of Engineering, The University of Tokyo, \\ 7-3-1 Hongo, Bunkyo-ku, Tokyo 113-8656, Japan}
\affiliation{Optical Quantum Computing Research Team, RIKEN Center for Quantum Computing, \\ 2-1 Hirosawa, Wako, Saitama 351-0198, Japan}
\author{Warit Asavanant}
\affiliation{Department of Applied Physics, School of Engineering, The University of Tokyo, \\ 7-3-1 Hongo, Bunkyo-ku, Tokyo 113-8656, Japan}
\affiliation{Optical Quantum Computing Research Team, RIKEN Center for Quantum Computing, \\ 2-1 Hirosawa, Wako, Saitama 351-0198, Japan}
\author{Jun-ichi Yoshikawa}
\affiliation{Optical Quantum Computing Research Team, RIKEN Center for Quantum Computing, \\ 2-1 Hirosawa, Wako, Saitama 351-0198, Japan}
\author{Mamoru Endo}
\affiliation{Department of Applied Physics, School of Engineering, The University of Tokyo, \\ 7-3-1 Hongo, Bunkyo-ku, Tokyo 113-8656, Japan}
\affiliation{Optical Quantum Computing Research Team, RIKEN Center for Quantum Computing, \\ 2-1 Hirosawa, Wako, Saitama 351-0198, Japan}
\author{Akira Furusawa}
\email{akiraf@ap.t.u-tokyo.ac.jp}
\affiliation{Department of Applied Physics, School of Engineering, The University of Tokyo, \\ 7-3-1 Hongo, Bunkyo-ku, Tokyo 113-8656, Japan}
\affiliation{Optical Quantum Computing Research Team, RIKEN Center for Quantum Computing, \\ 2-1 Hirosawa, Wako, Saitama 351-0198, Japan}


\date{\today}

\begin{abstract}
Non-Gaussian states of light, which are essential in fault-tolerant and universal optical quantum computation, are typically generated by a heralding scheme using photon detectors. Recently, it is theoretically shown that the large timing jitter of the photon detectors deteriorates the purity of the generated non-Gaussian states [T. Sonoyama, \textit{et al}., Phys. Rev. A \textbf{105}, 043714 (2022)]. In this study, we generate non-Gaussian states with Wigner negativity by time-gated photon detection. We use a fast optical switch for time gating to effectively improve the timing jitter of a photon-number-resolving detector based on transition edge sensor from 50 ns to 10 ns. As a result, we generate non-Gaussian states with Wigner negativity of $-0.011\pm 0.004$, which cannot be observed without the time-gated photon detection method. These results confirm the effect of the timing jitter on non-Gaussian state generation experimentally for the first time and provide the promising method of high-purity non-Gaussian state generation.
\end{abstract}


\maketitle

\begin{figure}
\includegraphics[width=\columnwidth]{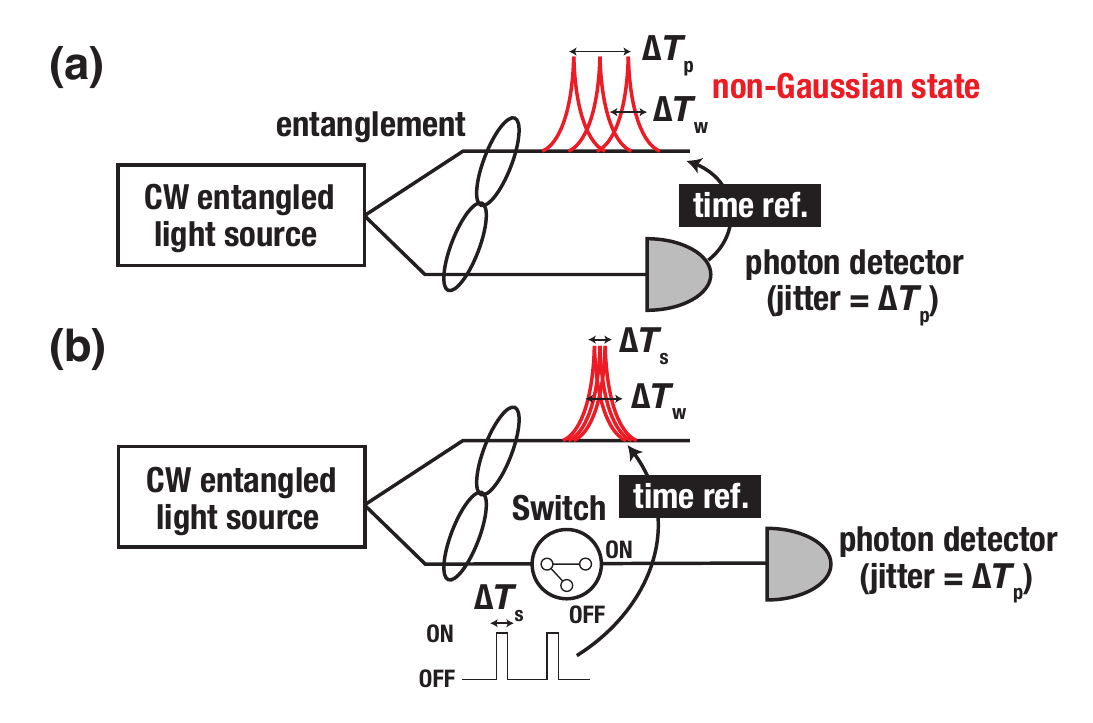}
\caption{Schematic diagram of non-Gaussian-state preparation by a heralding scheme using a continuous-wave (CW) entangled light source. (a) Conventional photon detection method. The detector's output is used as a time reference for determining state preparation timing. Thus, the temporal resolution is determined by the original timing jitter of photon detector $\Delta T_p$. Here, we note that $\Delta T_{\rm w}$ is the temporal width of the each wave packet. (b) Time-gated photon detection method. An optical switch is put just before the photon detector and operated in a short time window. By using the driving signal as a time reference, the temporal resolution is determined by the time window $\Delta T_s$, which becomes the effective timing jitter.}
\label{Fig:nonGauss}
\end{figure}

\textit{Introduction.-}
Non-Gaussian states of light are important resources in various fields such as quantum metrology \cite{PhysRevA.78.063828,PhysRevLett.104.103602}, quantum communication \cite{PhysRevA.90.062316,PhysRevA.95.062309}, quantum computation \cite{PhysRevLett.82.1784, Andersen:2015aa, PhysRevA.64.012310, PhysRevA.59.2631,PhysRevLett.111.120501,PhysRevX.6.031006,PhysRevA.94.042332,PhysRevX.10.031050,PhysRevX.10.011058}. Especially in the field of continuous-variable quantum information processing, quantum computation with Gaussian states and Gaussian operations have already been demonstrated \cite{Warit-cluster, Mikkel-cluster, BarameeQC, Larsen-2modeGaussian}, and the remaining issues are quantum computation with Non-Gaussian states and non-Gaussian operations, which are necessary for fault-tolerance and universality.\\
\indent In the optical systems, non-Gaussian states are typically generated by a heralding scheme using entangled resource and photon detectors \cite{PhysRevLett.56.58, PhysRevA.55.3184, pulsed-cat, Cat-continuous} as in Fig.\ref{Fig:nonGauss}. However, the purity of non-Gaussian states generated experimentally, which is crucial for their applications, is still limited because of the performances of photon detectors such as detection efficiency and photon-number-resolving ability. Recently, it has become theoretically clear that the timing jitter of the photon detector also affects the purity of the generated states when continuous-wave (CW) light is used \cite{PhysRevA.105.043714}. In particular, when the magnitude of the timing jitter is large and non-negligible relative to the wave packet width of the generated state, the timing of the state generation is unclear as shown in Fig.\ref{Fig:nonGauss} (a) and the purity is degraded. Most of the previous studies used single-photon detectors \cite{Cova:1981aa, Korzh:2020aa} with low timing jitter of less than 1 nanosecond to generate states on wave packets of several tens of nanoseconds \cite{Yukawa-fock,Wakui-cat,Warit-cat, Takase:22}. Thus, the effect of timing jitter was low enough to ignore. On the other hand, the use of photon-number-resolving detectors (PNRDs) is needed for generating more complex non-Gaussian states. The timing jitter of the typical PNRD based on Transition Edge Sensors (TESs) \cite{Fukuda:11} is about 50 ns, which is non-negligible compared to the typical wave packet width of several tens of nanoseconds. Furthermore, considering its application in quantum computation, the wave packet width is expected to become even shorter for ultrafast quantum computation. This is because the wave packet width is related to the clock frequency when time-domain multiplexing scheme is used, which is one of the promising methods of optical quantum computation \cite{Warit-cluster, Mikkel-cluster, BarameeQC, Larsen-2modeGaussian}. Therefore, the relative magnitude of timing jitter to wave packet width is expected to become larger and it has become necessary to deal with this problem of timing jitter in non-Gaussian state generation.\\
\indent Here, we introduce time-gated photon detection using a high-speed optical switch to improve the timing jitter in non-Gaussian state preparation. In conventional photon detection, the output signal of the detector is used as a timing reference of the state generation as shown in Fig.\ref{Fig:nonGauss} (a). In this case, the temporal resolution is determined by the detector's timing jitter $\Delta T_{p}$. On the other hand, in time-gated photon detection, an optical switch is placed in front of the detector as shown in Fig.\ref{Fig:nonGauss} (b), which operates only for a short time $\Delta T_{s}$ to limit the photon detection time. By using the ON/OFF signal of the optical switch as the reference signal for the state generation time, the temporal resolution is determined by the time window $\Delta T_{s}$ of the optical switch, which becomes the effective timing jitter. Thus, the timing jitter can be improved when the time window $\Delta T_{s}$ of the optical switch is shorter than the original timing jitter $\Delta T_{p}$ of the detector.\\
\indent In this study, a Pockels cell and a polarizing beam splitter (PBS) are used as an optical switch, and the effective timing jitter of a TES is controlled by the time window of the optical switch. As a result, the timing jitter of the TES is improved from 50 ns to 10 ns, and a Schr\"{o}dinger cat state, one of non-Gaussian states, with Wigner negativity of $-0.011\pm 0.004$ is successfully generated. This is in contrast with the case without time-gated photon detection, where the negative value of the Wigner function cannot be confirmed. To our knowledge, this is the first successful generation of Schr\"{o}dinger cat states using TESs and CW light. These results show the effect of timing jitter on the generated states experimentally and provide a promising method for generating high-purity non-Gaussian states.\\

\textit{Temporal mode functions of optical quantum states.-}
The optical quantum states we are dealing with exist in a wave packet. By defining the temporal mode function $f(t)$ as the function that represents the temporal shape of the wave packet, the complex amplitude of the wave packet $\hat{a}_{f}$ is expressed using the amplitude $\hat{a}(t)$ of the electromagnetic field at time $t$ as follows,
\begin{gather}
\hat{a}_f = \int_{-\infty}^{\infty} dt f(t) \hat{a}(t),
\end{gather}
where $f(t)$ satisfies $\int_{-\infty}^{\infty} dt \abs{f(t)}^2 = 1$. Here we consider rotating frame so that $f(t)$ does not include the oscillation components of the carrier frequency. Since the complex amplitude $\hat{a}_{f}$ is not an observable, the quadratures corresponding to the real and imaginary components $\hat{x}_f = \frac{\hat{a}_{f} + \hat{a}^{\dagger}_{f}}{\sqrt{2}}, \hat{p}_f = \frac{\hat{a}_{f} - \hat{a}^{\dagger}_{f}}{\sqrt{2}i}$ are used to describe quantum states. The quadratures satisfy the commutation relation $[\hat{x}_f,\hat{p}_f] = i$, where $\hbar$ is set to 1. In this paper, we denote the density matrix and the state vector of the quantum state excited on the temporal mode $f(t)$ as $\hat{\rho}_{f}$ and $\ket{\psi}_{f}$.\\

\textit{Heralded generation of Shr\"{o}dinger cat states.-} In this experiment, a Schr\"{o}dinger cat state, known as a fundamental element of optical quantum information processing, is chosen as a non-Gaussian state to be generated. This Schr\"{o}dinger cat state is a superposition of a coherent state, i.e., $\ket{\rm cat} = \frac{1}{N_{\psi,\alpha}}(\ket{\alpha} + e^{i\psi}\ket{-\alpha})$, where $\psi$ is the relative phase and $N_{\psi,\alpha}$ is a normalization constant. Such a cat state is generated by subtracting a photon from the squeezed light, limiting the frequency region to be measured with a frequency filter and then detecting the photons \cite{Cat-continuous, pulsed-cat, Wakui-cat, Warit-cat, Takase:22, Namekata:2010aa, PhysRevA.82.031802}. \\
\indent Here, we discuss the effect of the timing jitter of photon detection on the generated quantum state quantitatively. The theoretical analysis for the case of single-photon-state generation has been presented in \cite{PhysRevA.105.043714}, so in this paper we extend it to the case of Schr\"{o}dinger-cat-state generation as well. First, if the timing jitter is sufficiently low, it is known that the generated quantum state is expressed as follows \cite{Warit-cat, Takase:22},
\begin{gather}
\hat{a}_{N(g)} \hat{S}_{r} \ket{0} \propto \hat{S}_{r}\hat{a}^{\dagger}_{N(g*r)}\ket{0},
\end{gather}
where $N(\cdot)$ is a normalizing operation, $g(t)$ is a time-reversed function of the frequency filter's impulse response, $r(t)$ is the time correlation function of the squeezed light and $\hat{S}$ is a squeezing operator. By defining $f(t)$ as $g*r(t)$, this quantum state can be approximated as $\ket{\rm cat}_{f}$. On the other hand, if the timing jitter is poor and its effect cannot be ignored, the quantum state generated is a mixture of cat states excited on wave packets with different times, as shown in Fig.\ref{Fig:nonGauss} (a). Assuming the distribution function of timing jitter as $j(t)$, the generated state $\hat{\rho}$ can be expressed as follows,
\begin{gather}
  \hat{\rho} \propto \int_{-\infty}^{\infty} dt' j(t') \ket{\rm cat}_{f(t-t')} \bra{\rm cat}_{f(t-t')}.
\label{eq:jt}
\end{gather}
Such a quantum state $\hat{\rho}$ is a multimode state in multiple wave packets, but what we want to generate is a single-mode Schr\"{o}dinger cat state that can be used for quantum computation. We consider therefore choosing a temporal mode $f_1(t)$ such that the quantum state on $f_1(t)$ is closest to the cat state. This $f_1(t)$ can be obtained by a principal component analysis (PCA) method \cite{PhysRevA.105.043714}. Then, the generated state $\hat{\rho}_{f_1}$ in the temporal mode $f_1(t)$ can be expressed as follows using the cat state $\ket{\rm cat}$ and the squeezed state $\ket{\rm sqz}$:
\begin{gather}
\hat{\rho}_{f_1} =  \lambda_{1} \ket{\rm cat}\bra{\rm cat}_{f_1} + (1-\lambda_1) \ket{\rm sqz} \bra{\rm sqz}_{f_1},
\label{eq:mixedstate}
\end{gather}
where the parameter $\lambda_1$ and the temporal mode function $f_1(t)$ depend on the ratio of the timing jitter to the time width $\Delta T_{\rm w}$ of the original wave packet $f(t)$. When the timing jitter relative to $\Delta T_{\rm w}$ is low enough, $\lambda_1$ approaches 1 and $f_1(t)$ approaches $g*r(t)$. On the other hand, when the timing jitter becomes worse, $\lambda_1$ becomes smaller and $f_1(t)$ becomes wider in time. Thus, the generated quantum state becomes a mixture of the cat state (non-Gaussian state), and the squeezed state (Gaussian state).\\
\indent The timing jitter dependence on the generated quantum states can be confirmed experimentally. First, the optimal time mode $f_1(t)$ can be obtained by PCA method, and the timing jitter's effect is confirmed by seeing how the shape of $f_1(t)$ changes with the magnitude of the timing jitter. In addition, the purity of the quantum state, which is related to $\lambda_1$ in Eq.\ref{eq:mixedstate}, can be confirmed from the photon number distribution of the generated quantum states and the value of the Wigner function near the origin. This is because the squeezed light has an even-photon nature and the cat state generated by the single-photon subtraction in this experiment has an odd-photon nature \cite{PhysRevA.55.3184}, and the value at the origin of the Wigner function is $-\frac{1}{\pi}$ for a quantum state with odd-photon nature and $\frac{1}{\pi}$ for a quantum state with even-photon nature \cite{PhysRevA.15.449,Nehra:19}. In this paper, these properties are used to evaluate the generated quantum states.\\ 

\textit{Experimental setup.-}
\begin{figure}
\includegraphics[width=\columnwidth]{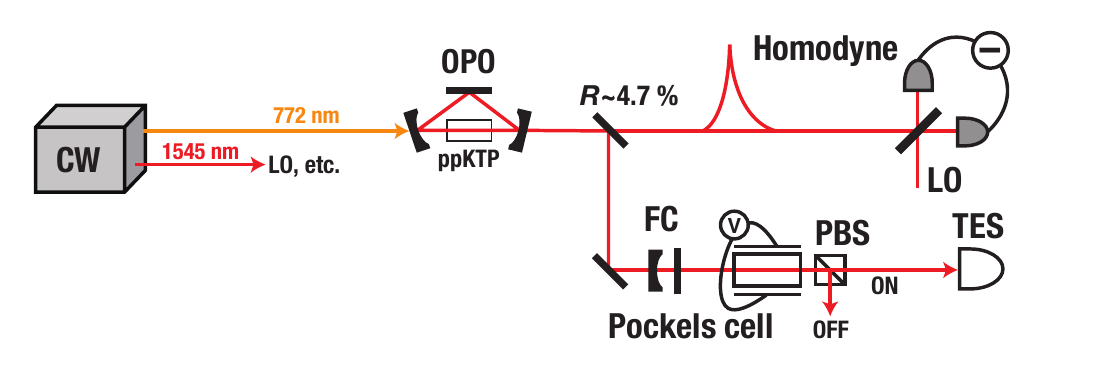}
\caption{Simple schematic diagram of the experimental setup for Shr\"{o}dinger-cat-state generation with time-gated photon detection. An optical switch consists of a Pockels cell and a Polarizing Beam Splitter (PBS). OPO: Optical Parametric Oscillator (HWHM =  58.4 MHz), FC: Filter Cavity (HWHM = 8 MHz), ppKTP: periodically poled ${\rm KTiOPO_{4}}$, LO: Local Oscillator.}
\label{Fig:Experiment}
\end{figure}
In this experiment, Schr\"{o}dinger cat states are generated by photon subtraction both with and without the optical switch for comparison. In particular, five patterns are performed with the optical switch (time window of 10, 30, 50, and 70 ns) and without the optical switch (the original timing jitter of the TES is $\Delta T_{\rm p} = $  58 ns, which is estimated experimentally). \\
\indent  The experimental setup is shown in Fig.\ref{Fig:Experiment}, where the light source is a CW laser at 1545.32 nm and the squeezed light is generated by an Optical Parametric Oscillator (OPO) whose half-width-at-half-maximum (HWHM) is 58 MHz. A small portion (4.7\%) of the generated squeezed light is tapped and a single photon is detected by the TES after a frequency filtering whose HWHM is 8 MHz. These parameters determine the wave packet width $\Delta T_{\rm w}$ of the generated state when the timing jitter of photon detection can be neglected. The theoretically expected value is about $\Delta T_{\rm w} = $22 ns. Here, a Pockels cell and a PBS are used as an optical switch and the minimum time window of this optical switch is 10 ns, which is small compared to the temporal width of the wave packet $\Delta T_{\rm w}$. The Pockels cell used in this study is ${\rm RTiOPO}_{4}$, which has the extinction ratio of more than 30 dB and the transmittance of more than 95 \%. The Pockels cell driver, which drives the voltage applied to the Pockels cell, has a rise time of about 3.5 ns and a maximum repetition rate of 1.2 MHz. \\
\indent For verification, quadratures of the generated quantum states at time $t$, $\hat{x}_{\theta}(t) = \hat{x}(t)\cos\theta +  \hat{p}(t)\sin\theta,\,(\theta = 0,\frac{1}{6}\pi,\frac{1}{3}\pi,\frac{1}{2}\pi,\frac{2}{3}\pi,\frac{5}{6}\pi)$ are obtained by homodyne measurements, from which the temporal mode functions and Wigner functions of the generated states are reconstructed. The count rates in this experiment are 17 ${\rm s}^{-1}$ (with the switch, 10 ns), 52 ${\rm s}^{-1}$ (30 ns), 79 ${\rm s}^{-1}$ (50 ns), 108 ${\rm s}^{-1}$ (70 ns) and 3900 ${\rm s}^{-1}$ (without the switch), respectively. \\

\textit{Evaluation.-}
\begin{figure}
\includegraphics[width=\columnwidth]{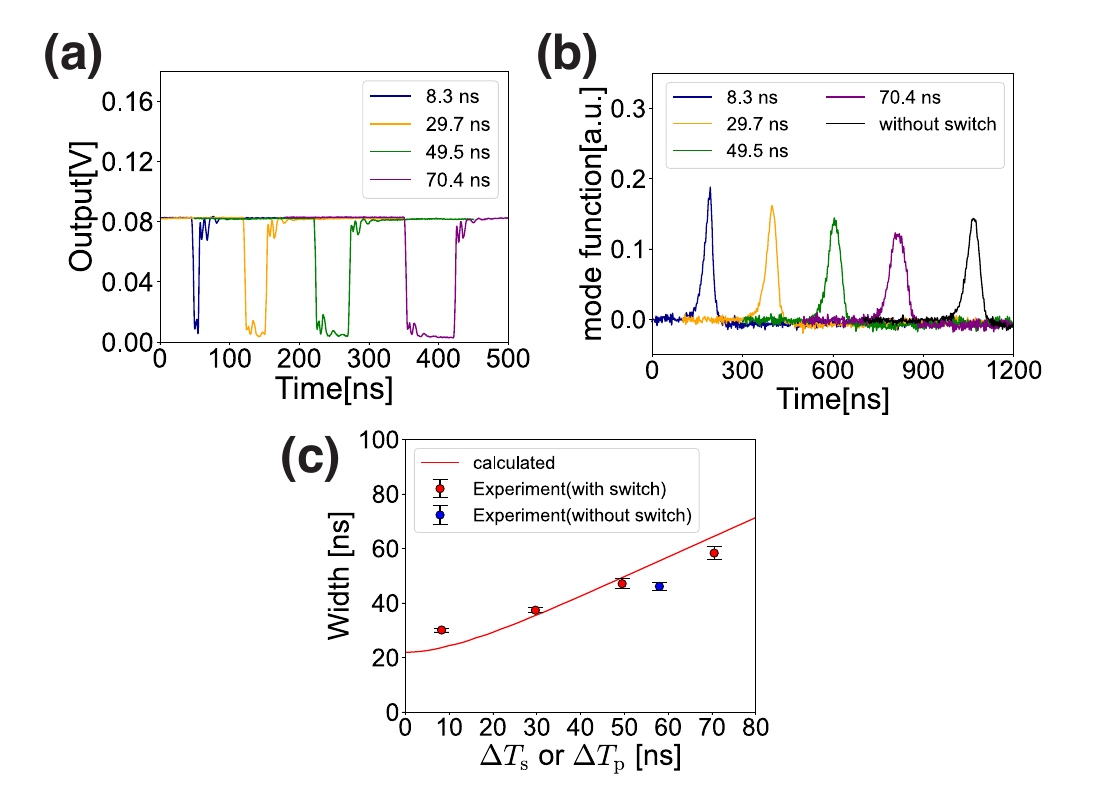}
\caption{(a) Response of the optical switch to an input of classical light. To improve visibility, each signal has different offsets in the horizontal axis. (b) The estimated temporal mode functions $f_1(t)$ of the generated quantum states. Here, when the optical switch is not used, the timing jitter is TES's original jitter $\Delta T_{\rm p} = 58$ ns. As in (a), each signal has different offsets. (c) The plots of the time width (FWHM) of the estimated temporal mode functions and the time width (FWHM) of the theoretically calculated temporal mode functions.  $\Delta T_{\rm w}$ corresponds to the time width when the timing jitter $\Delta T_{\rm p}$ or $\Delta T_{\rm s}$ is 0. In the numerical calculations, the bandwidth of the OPO (HWHM), the bandwidth of the Filter cavity (HWHM) are assumed to be 58.4 MHz, 8 MHz, respectively. In addition, the distribution function of the timing jitter $j(t)$ is assumed to be a rectangular function. Here, the bootstrap method is used to obtain the error of the temporal mode function estimation.}
\label{Fig:temporalmodes}
\end{figure}
\begin{figure*}
\includegraphics[width=2\columnwidth]{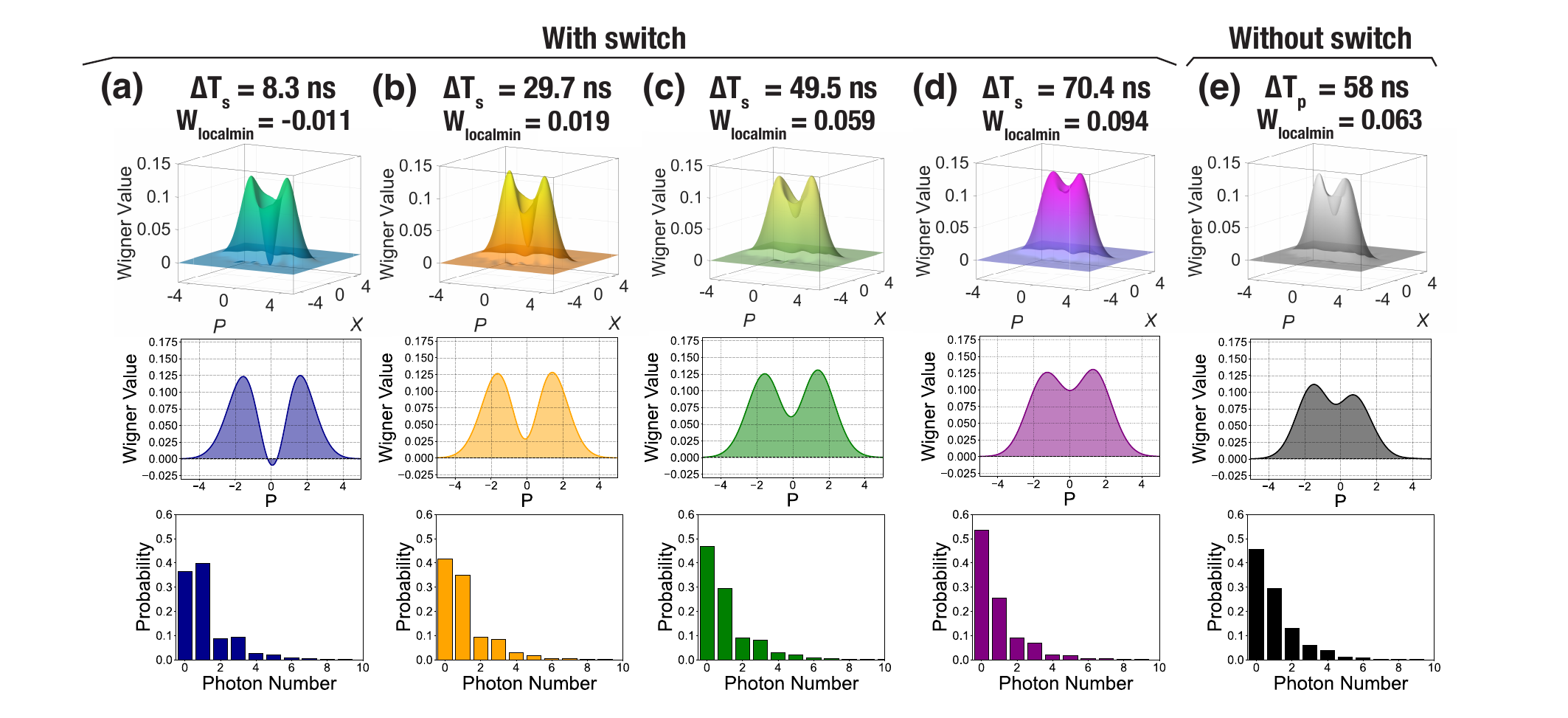}
\caption{(a)-(e) The reconstructed Wigner functions, their cross sections at $X=0$, and photon number distributions with the optical switch ($\Delta T_{\rm s} = $ 8.3, 29.7, 49.5 and 70.4 ns) and without the optical switch ($\Delta T_{\rm p}$ = 58 ns). The local minimums around the origin of the Wigner functions are $-0.011 \pm 0.004, 0.019 \pm 0.004, 0.059 \pm 0.004, 0.094 \pm 0.003$ ($\Delta T_{\rm s} = $ 8.3, 29.7, 49.5 and 70.4 ns), $0.063 \pm 0.003$ ($\Delta T_{\rm p}$ = 58 ns), respectively. The bootstrap method is also used to obtain the estimation error. We note that the minimum value near the origin and the value at the origin differ slightly in these reconstructed Wigner functions.}
\label{Fig:Wigners}
\end{figure*}
The first step is to check the operation of the optical switch. Classical coherent light is injected into the optical switch, and a fast photodetector is placed immediately after the PBS to check its response. Figure \ref{Fig:temporalmodes} (a) shows the response of the optical switch when the time window is set to 10, 30, 50, and 70 ns, respectively. The actual time widths are $\Delta T_{\rm s} = 8.3, 29.7, 49.5$ and $70.4$ ns, which are consistent with the set values.\\
\indent Next, the temporal mode functions of the generated quantum states are reconstructed. The temporal mode functions are obtained by PCA of the covariance matrix of the quadratures, taking advantage of the fact that the quadrature's variance of the generated quantum state is larger than the background quantum state \cite{PhysRevLett.109.033601,PhysRevLett.111.213602}. Figure \ref{Fig:temporalmodes} (b) shows the estimated temporal mode functions $f_1(t)$ of the generated states with and without the optical switch. This $f_1(t)$ corresponds to the first principal component obtained by PCA. Theoretically, as the timing jitter becomes worse, the width of the estimated temporal mode $f_1(t)$ should widen in time and its peak value should become smaller, and such a tendency is confirmed in Fig.\ref{Fig:temporalmodes}. In addition, Fig.\ref{Fig:temporalmodes} (c) compares experimental and numerically calculated theoretical values of the full-width-at-half-maximum (FWHM) of the temporal mode function $f_1(t)$ with respect to the timing jitter. The results with the optical switch are generally consistent with the theoretical line, but the result without the optical switch seems to deviate a little from the theoretical line. This may be due to the fact that the shape of distribution function $j(t)$ in Eq.\ref{eq:jt} is not the same as in the case of using an optical switch, or the output signal of the TES changed during the experiment and the timing jitter may have differed from the estimated one. From these results, we confirm the improvement of the timing jitter and the change of the temporal mode function by time-gated photon detection. \\
\indent Finally, the quadratures in the temporal mode $f_1(t)$ are calculated numerically as $\hat{x}_{f_1,\theta} = \int dt f_1(t) \hat{x}_{\theta}(t).$ Then, the density matrices and the Wigner functions are reconstructed from the quadrature data by quantum state tomography \cite{RevModPhys.81.299} without any loss correction. The reconstructed Wigner functions, their cross sections at $X=0$, and the photon number distributions are shown in Fig.\ref{Fig:Wigners}. As shown in Fig.\ref{Fig:Wigners} (a), the Shr\"{o}dinger cat state with a negative value of $-0.011 \pm 0.004$ near the origin is successfully generated when the time window of the optical switch is the shortest ($\Delta T_{\rm s} = 8.3$ ns). The photon number distribution also confirms the odd-photon nature of the generated state, which is caused by subtracting a single photon from the squeezed state with even-photon nature. This shows that we succeed in generating a Shr\"{o}dinger cat state. Figure \ref{Fig:Wigners} (b)-(d) show that as the optical switch time window $\Delta T_{\rm s}$ becomes wider (29.7 ns, 49.5 ns and 70.4 ns), the value of Wigner function near the origin gradually increases. Furthermore, Fig.\ref{Fig:Wigners} (e) shows the Wigner function of the generated state without the optical switch ($\Delta T_{\rm p} = 58$ ns) and the negative value near the origin cannot be observed here. From these results, we confirm that the improvement of timing jitter leads to the improvement of non-Gaussian states generation, as expected theoretically.\\

\textit{Conclusion.-} In this study, we introduce time-gated photon detection using a high-speed optical switch for non-Gaussian state generation. As a result, the effect of the timing jitter of the detector on the non-Gaussian state generation is experimentally clarified for the first time, and it is confirmed that the purity deteriorates when the timing jitter is non-negligibly large compared to the wave packet width of the generated state as expected theoretically. In addition, we improve the timing jitter of the PNRD-TES from 50 ns to 10 ns using this method, and succeed in generating a Schr\"{o}dinger cat state with a negative Wigner function value of $-0.011$. To the best of our knowledge, this is the first successful generation of Schr\"{o}dinger cat state using TES and a CW light source.\\
\indent This time-gated photon detection method does not sacrifice the other performances of the detector TES such as detection efficiency and photon-number-resolving capability. Thus, we expect more complex non-Gaussian state generation is possible by using this time-gated scheme in multi-photon detection regime. Furthermore, by using further faster optical switches using non-linear optical effect as in \cite{England:21,Kupchak:19}, the timing jitter of a photon detector could be further improved to the subpicosecond regime. This means that this scheme can be used to generate high-purity non-Gaussian states in picosecond wave packet. This technology is therefore expected to become the key technology for realizing ultrafast, fault-tolerant and universal optical quantum computation with GHz to THz clock frequency.\\

The authors acknowledge supports from UTokyo Foundation and donations from Nichia Corporation of Japan. W.A. and M.E. acknowledge supports from Research Foundation for OptoScience and Technology. T.S., K.Takahashi and B.C acknowledge financial supports from The Forefront Physics and Mathematics Program to Drive Transformation (FoPM). The authors would like to thank Mr. Takahiro Mitani for careful proofreading of the manuscript. This work was partly supported by Japan Society for the Promotion of Science KAKENHI (18H05207, 20J10844, 20K15187, 22K20351), Japan Science and Technology Agency Moonshot Research and Development (JPMJMS2064), PRESTO (JPMJPR2254), and CREST(JPMJCR17N4). 

\bibliography{paper1.bib}

\end{document}